\documentclass{elsarticle}
\usepackage{times}
\usepackage{palatino}
\usepackage{amsmath}
\usepackage{amsfonts}
\usepackage{amssymb}
\usepackage{graphicx}
\usepackage{graphicx,color}
\usepackage{alltt}
\usepackage{mathrsfs}
\usepackage{epsfig}
\usepackage{subfigure}
\usepackage{verbatim}
\begin{document}
\biboptions{longnamesfirst} 
%\preprint{}
\title{ Arbitrary  bending of  optical
 solitonic beam regulated  by  
boundary excitations  in
 a doped resonant medium}

\author[ak]{Anjan Kundu} \ead{anjan.kundu@saha.ac.in} \address[ak]{Theory
Division, Saha Institute of Nuclear Physics, 1/AF, Bidhannagar, Kolkata
700064, India\\ {Temporary address}: International Centre for Theoretical
Physics, Trieste, Italy }

\author[tn]{Tapan Naskar} \ead{tapan.naskar@cbs.ac.in} \address[tn]{
 Centre for Excellence in Basic Sciences,  University of
Mumbai,
Vidhyanagari Campus,
Mumbai - 400098, India}
\date{\today}

\begin{abstract}
 Bending of a shape-invariant optical beam is achieved so far along
parabolic or circular curves.
  Borrowing  ideas used in nonlinear optical communication, we propose such
a bending along any preassigned curve or surface,
   controlled by the boundary population inversion of atoms in an Erbium
 doped   medium.  The optical beam generated in a nonlinear Kerr medium and
  transmitted through a doped resonant  medium as an accelerating soliton,
 predicted here, should be realizable experimentally and applicable to
 nonlinear events in other areas like plasma or ocean wave.
  \end{abstract}
%76 words
%\pacs{
%%%42.65.Tg,
%optical soliton, nonlinear waves 
%%%02.30.lk,
%integrable system
%%% 05.45.Yv,
%soliton, 
%42.65.-k,
%nonlin optics
%42.81.Dp,   
%prop soliton
%42.50.Md
%self-induced transparency 
%%%42.65.Sf}
%dynamics Nlin optics
\begin{keyword}
bending of light beam, accelerating soliton, nonlinear optics, integrable system
\end{keyword}
\maketitle
%\section{Introduction}\label{intro}
\section{Introduction}
\subsection{Bending of light beam} It is our  long-standing dream, that it would be possible some day 
 to produce a localized light beam in a medium, that could be bent to any
preassigned curve maintaining its
  shape throughout the propagation.  Such
a self-accelerating beam would be able to self-bend around an obstacle,
  making it virtually invisible.  However, the first step in achieving
 the bending of an optical   beam through
theoretical and experimental studies  was taken  quite 
 late  \cite{1,2}, inspired by a pioneering work of Berry et al related to
 accelerating free
quantum particles \cite{berry}.  The 
 beam,  typically an Airy function solution of linear Schr\"odinger
equation,  could preserve   its  parabolic form over a finite
distance.
 Subsequently, acceleration along an arbitrary curve was achieved, though at
  the cost of non-preservation of the shape
 \cite{shapeNP}.
  Meanwhile  bending of propagating solitons has been achieved in other
fields, e.g. ring accelerator for matter-wave solitons \cite{carpentier07},
 in chirped photonic lattices \cite{kartashov05},  self-bending of soliton
in optical lattices \cite{kartashov04} etc.
 
Bending of an optical beam along a   circular path  
is reported  recently for a shape-sustaining optical
 beam on a plane,  breathing periodically in time, 
as a Bessel function related  
  exact solution to the linear Maxwell equation in vacuum 
 \cite{circle}.
 
A parallel development has   taken place   
 involving nonlinear media,  
in which an  Airy-like beam  solution  was found  to
 the nonlinear   Schr\"odinger (NLS) equation in a Kerr medium
 \cite{NLtheor,NLth-exp}.  Negating the prevailing
scepticism, that symmetric nonlinearities
 can not produce self-accelerating soliton,  a shape-invariant
  static beam 
  oriented along a parabolic curve on a plane was obtained and 
 generalized  also  to 3D, which subsequently
 have been verified in real experiments \cite{NLexp,NLth-exp}.

Nevertheless, in all these  exciting achievements,  
  the bending of a  shape-preserving beam is restricted 
  basically to   two specific 
 curves: circular or parabolic, linked to the Airy or Bessel function
  solutions \cite{1}-\cite{berry},\cite{circle}-\cite{NLth-exp}.  
This situation however falls
way short of our expectation of having  a stable
  optical beam that could  bend along any desired curve, regulated by an
arrangement  in the  medium and  turn  any object invisible by bending
 around  it. 
We offer here  a breakthrough by proposing a scenario close to the above 
 expectation, where a localized 2D or 3D optical
  beam could  be accelerated arbitrarily, resulting to its bending 
along an arbitrary  curved path. The nature of the curve 
 can be controlled by the boundary
 configuration of the population inversion of atoms in a doped resonant 
 medium, through which the shape-invariant beam passes.  Therefore, the
optical beam can turn around an object of arbitrary shape without hitting
it.  The idea of our construction for the  bending of an optical beam follows a
path, fundamentally different from those adopted so far, as briefed above,
 and is inspired by a series of earlier studies in the nonlinear optical
 communication \cite{nlssolit}-\cite{AIPad11}. Though our proposal is based
on theoretical models we hope its feasibility through future experiments. 

\subsection{Solitonic optical communication}
 %%-------------
 Soliton based optical communication rests
  on the principle that, when the intensity of the optical pulses crosses a threshold value the fiber medium behaves nonlinearly, since the refractive index becomes intensity dependent. Moreover, the self-phase modulation (SPM)
 due to Kerr nonlinearity acts in opposition to the dispersion in the medium 
 at the anomalous dispersion regime. At proper choice of the parametric set
 the effect of dispersion could be at perfect balance with that of the nonlinear SPM
 creating a condition for the stable solitonic pulse propagation of the electric field
 $E(t,z)$, with its dynamics
 %---------------_____________
 governed by the 
 NLS equation
\begin{eqnarray} iE_z+E_{tt}+2|E|^2E=0. \label{1DNLS} \end{eqnarray}
 The well known and well studied NLS equation (\ref{1DNLS})
can be derived from the Maxwell equation in a nonlinear medium \cite{agarwal}.
 Note that, fiber communication through an electric field,
  transmitted  in a Kerr medium as a
  stable solitonic  pulse  moving
 with a constant  velocity,
 was proposed long ego \cite{nlssolit}.

 Another proposal,  which gained popularity for improved transmission of
 nonlinear optical pulses, is due to self-induced transparency (SIT) produced
 by the coherent response of a medium to ultra short pulses
 \cite{SIT}.
 %%%%%%%%%%----------------------------------
  The medium is doped  usually with Erbium ions, since 
  this rare earth material prominently exhibits an effective two-level
 energy spectrum (with $j=\frac {15} 2 $, as the ground state and $j=\frac
 {13} 2 $, as the excited state), though with a smearing of the levels due
 to the Stark effect.  Moreover,
  the energy difference between these    two levels is nearly equal to that
of the wavelength at which soliton pulses are propagated.  Therefore the
resonant interaction makes the two-level medium optically transparent to
that wavelength, inducing a SIT process.
 %%%%%%%%%%% ------------ 
 %oooooooooooo
  The interacting fields
involve   polarization $p$ in the  resonant  medium 
 induced by 
the propagating optical field $E$ 
and   population inversion (PI) $N$   of an ensemble of  two-level atoms.
%%%%%%%-------------
The resonant interaction these fields makes the two-level medium transparent to the
wavelength at which the optical pulses propagate.  Amplification and
long-distance soliton transmission is now successfully achieved  for
practical purposes using  distributed Er-doped fibers
\cite{rottwitt93,suzuki89,wen94}.
 %%%%%%%%%%--------------------------------------
 The  dynamics of the fields is  described by the SIT equations
\begin{eqnarray}iE_{z}=2p,\  ip_t=2(NE-\omega_0p), \ iN_t=E^*p-p^*E,
 \label{SIT}
\end{eqnarray}
which 
are reduced from the Maxwell-Bloch (MB) equations 
through  an 
ensemble average together with 
 assumptions of a sharp resonance and  homogeneous broadening
\cite{SIT,maimistov,nakazawa}. The SIT equations  represent also  an integrable
system generating solitonic pulses moving with a constant velocity \cite{SIT}.

 A subsequent  important development is to combine these two models by
suitably fabricating the optical fibers for coherent transmission of optical
pulses.  When a two-level resonant medium such as Er is doped with the core
of the optical fiber the propagation of optical soliton is described by the
MB equations (\ref{SIT}), whereas the silica material of the fiber induces
the solitonic propagation due to NLS equation (\ref{1DNLS}).  Therefore the
wave propagation can have both these effects due to silica and doped medium,
with Er-impurities accounting for the SIT effect.  The dynamics of the
combined system, with wave propagation in Er-doped optical fibers is
governed by the coupled NLS-SIT equations
\begin{eqnarray}iE_{z}+E_{tt}+2|E|^2E=2p,
% \label{NLSp}\end{eqnarray} 
 % coupled to the stationary SIT equation
%\begin{eqnarray} 
\nonumber \\
ip_t=2(NE-\omega_0p), \ iN_t=E^*p-p^*E.
\label{SITt}
\end{eqnarray}
 In this  NLS-SIT system an optical soliton,  
 created in a Kerr medium, is  transmitted through an Erbium doped resonant
medium inducing polarization
  and excitation of the two-level  dopant atoms, which has been
%_------------ 
theoretically predicted first by Maimistov and Manyakin \cite{maimistov}
 and since then been experimentally verified
  \cite{nakazawa}, extended further \cite{porsezian,AIPad11} and   implemented for
practical usage \cite{bhadraBOOK13,rottwitt93,suzuki89,wen94}.  Further
details on this combined system can be found in a recent monograph
\cite{bhadraBOOK13}.
%__________________
Since the ensemble of dopant atoms in the doped medium act as an effective
two-level system with the energy gap between the levels being close to that
of the
 wavelength of the propagating solitonic pulses,
the  resonant interaction   induces  a balance between the optical
 absorption
and emission,     sustaining a stable  shape-preserving  pulse propagation.
Remarkably, this coupled  system    turns out also  
 to be  integrable  admitting exact  soliton solutions
for all  fields  propagating with the same constant velocity.
\subsection{The present aim }
Our intention is to borrow the idea of this combined system, already in use 
in nonlinear optical communication and apply it  in the present  context
 of the optical beam bending on a 2D plane, by  using the same 
 governing equations (\ref{SITt}), though adopting suitably for the current
purpose.
%%-------------------------
To achieve  this we shift  to a moving frame: $ x=\frac 1 v (z+vt) $
and map the  fields as \begin{eqnarray}E(t,z) \to \tilde E(x,z), \
p(t,z) \to \tilde p(x,z),\ N(t,z) \to \tilde N(x,z) , \label{EpN} \end{eqnarray}
 Note that, since the NLSE and SIT    
equations are coupled, the solutions 
 created for all the fields $E,p,N $     
belonging to    these interacting equations move with the same group
velocity as well as the soliton velocity, which  
therefore makes it  possible to shift to  coordinates (x,z)
 in the moving frame with a fixed velocity,  simultaneously for all
the fields.
This  translates  the evolution in (\ref {SITt}) as
\begin{eqnarray}E_z(t,z)= \tilde E_z(x,z), \ E_{tt}(t,z)= \tilde E_{xx}(x,z),
p_t(t,z)=\tilde p_x(x,z),\ N_t(t,z)=\tilde N_x(x,z) , \label{EpNtx} \end{eqnarray}
due to the mutual independence of  coordinates $(t,z)$ as well as of  $(x,z) $,
 rewriting  equation (\ref{SITt}), 
suitable for optical communication,
 to the stationary equations convenient
 for the light beam propagation:
%%%%%-----------------------
\begin{eqnarray}iE_{z}+E_{xx}+2|E|^2E=2p, \label{NLSp}\end{eqnarray} 
  coupled to the stationary SIT equation
\begin{eqnarray} ip_x=2(NE-\omega_0p), \ iN_x=E^*p-p^*E,
\label{SITx}
\end{eqnarray}
involving induced polarization $ p$ 
   and the population inversion $N$ of the  resonant medium. In 
   (\ref{NLSp},\ref{SITx}) and their
   subsequent analysis we drop the $\ \tilde {} \ $ sign on the fields for
   convenience.
In the Erbium  doped  medium with 
 effective two-level atoms
 the 
 normalized wave functions may be given by  $ \nu$
  and $\tilde \nu$ for the ground and the excited states,  respectively
with  
 the optical field $ E(x,z)$ inducing polarization $ p(x,z)=\nu \tilde
 \nu^*$ in the resonant medium and
interacting with the PI 
$N(x,z)=|\tilde \nu|^2-|\nu|^2,$ of the  two-level  atoms.
  As a result the PI profile can change
within the range $\ -1 \ \leq N \leq \ +1 $ and
 all  interacting fields
satisfying 
 equations (\ref{NLSp}, \ref{SITx}) can
  {\it move} self-consistently in the form of stationary  solitonic pulses
  with a constant {\it space}-velocity.  In this set up therefore optical
  field can produce a localized straight beam without
 any bending.  However, at a closer look, it is important to identify that
the soliton
 velocity  in a SIT system is linked  to the PI of dopant atoms at the
  initial moment and its constancy is caused due to the conventional
 assumption of the initial PI  taken to be a constant: $N_0=1 $
 \cite{nakazawa,AIPad11}. 
 %%---------------
  Note that since we have moved from the variables
 $(t,z) $ to $(x,z) $ the time $t$ is replaced by the transverse dimension
 $x$ and as a consequence 
 the initial
condition in time $|t| \to \infty $ is to be replaced in the
 present context by a transverse boundary condition in $|x| \to \infty$, 
 %%%---------------
 as 
evident from the governing equations (\ref{NLSp}, \ref{SITx}).
 We will  see below what happens, when this boundary
  condition is relaxed and will use it in favor of the beam bending
scenario.

\section{ Arbitrary bending of a 2D beam} 
A  crucial observation for 
 coupled NLS-SIT system (\ref{NLSp},\ref{SITx}) is that,
 in place   of fixing the PI at the transverse boundaries: $N(x,z)_{x\to \pm
\infty}=N_0$ as a constant $\ N_0=1$, 
  we  may   choose   it to be an arbitrary function $N_0=N_0(z), $ varying
 along the optical axis.  Such a change may be made by exciting the dopant
 atoms at the boundaries not uniformly, but with variable intensities,
 irradiated suitably by laser pumping.  This alteration in preparing the
 setup, however, can bring a significant change in the character of the
 optical solitonic beam
propagating through the medium.
 Since the soliton velocity is linked to the boundary PI configuration in a
straightforward way, the velocity
   itself becomes a function of $z$ \cite{AIPad11}. 
 \subsection{Lax pair
formulation} For understanding the mechanism of creating an accelerating
solitonic beam we have to go deeper into the integrable structure of the
NLS-SIT system and look into its associated Lax pair $ ( U(\lambda), V(\lambda)) $ , which are given by a
deformation of the well known Lax pair $ (U_1(\lambda), V_2(\lambda))$ \cite{solit} 
associated with   the NLS equation as 
\begin{eqnarray}  U(\lambda)=U_1(\lambda), \  V(\lambda)= V_2(\lambda)+V_{-1}( \lambda
), \label{UV}
\end{eqnarray}
with the 	 addition of another matrix $ V_{-1}( \lambda)$, containing the SIT fields: $
 N,p,p^* $  
 and a negative power of the spectral parameter   $\lambda ^{-1} $
 \cite{nakazawa,porsezian,AIPad11}.  These Lax operators may be
 given  in the explicit form as 
 \begin{eqnarray} U_{1}(\lambda)= i(\lambda \sigma^3 +U^{(0)}), \ 
 \ U^{(0)}=
\left(\begin{array}{cc}
0 & E  \\
 E^* & 0 \\
 \end{array}
\right),
 \label{U1}\\
  V_2(\lambda)=2i (\lambda^2 \sigma^3 +\lambda U^{(0)})+\sigma^3(U^{(0)}_x-i{U^{(0)}}^2),
\label{V2}\end{eqnarray}
and 
\begin{eqnarray}
 V_{-1}(\lambda)= \frac 1 {\lambda- \omega_0} \left( 
\begin{array}{cc} N &p\\ p^* & -N \\ \end{array} 
\right)
\label{V-1}
\end{eqnarray}
It is not difficult to check that the flatness condition:
\begin{eqnarray}
 U_z (\lambda)- V_x(\lambda)+[ U (\lambda), V(\lambda)]=0
\label{flat}
\end{eqnarray}
 of the associated
Lax pair   (\ref{UV}) generates 
the  NLS-SIT equation (\ref{NLSp},\ref{SITx}) revealing the integrability of
the system, which admits an  exact soliton solution in the form
\begin{eqnarray}
E(x,z)&=& -2i{\rm sech} 2\xi (x,z) e^{2i\theta (x,z)}, \xi(x,z)=\eta( x-v(z)) ,
\nonumber \\ \theta(x,z)& =& -kx+\omega(z), 
\label{1DNLSpsolit}
\end{eqnarray} 
exhibiting  variable solitonic 
 velocity $ v(z)$ and the  modulation frequency $\omega(z),$
 linked to the asymptotic value  of the second Lax operator 
$V(\lambda) $ in (\ref{UV}) at $x\to\pm \infty$. Since the fields satisfy the boundary conditions 
  \begin{equation}
 E( x\to \pm \infty,z)= 0, \  p( x\to \pm \infty,z)=0, \  N( x\to \pm
\infty,z)=N_0(z), \label{BC00}\end{equation} where  the
boundary PI $ N_0(z) $ could be an arbitrary function of  $z$ along the optical axis, 
 the boundary value of the Lax operator $V(\lambda) $ can be given as
\begin{eqnarray} V(\lambda,x\to\pm \infty)=(V_2(\lambda)+V_{-1}( \lambda
))_{x\to\pm \infty} = (2i\lambda^2 + \frac 1 {\lambda-\omega_0
}N_0(z))\sigma^3.  \label{V-0} \end{eqnarray} Note that the first term in
the boundary value of the Lax operator (\ref{V-0} ), coming from the NLS system is a
constant, while the second term originating from the SIT system depends on
the boundary PI: $\ -1 \ \leq N_0(z) \leq \ +1, $ and could be
 an arbitrary function.
  
This possibility of achieving a $z$-dependent function at the boundaries in
place of the conventional choice of $N_0=1 ,$ gives rise to the variable
soliton velocity and modulation frequency (\ref{1DNLSpsolit}) as
\begin{eqnarray} v(z) =v_0z+v_m f(z), \ \ \omega(z)=\omega_0z+\omega_m f(z),
\label{Vz} \end{eqnarray} with the variable parts of the  velocity
and the frequency linked to   the PI
profile at the  transverse
 boundaries as $$ f(z)= \int_0 ^z dz' N_0(z').
$$
The constant parts of the velocity and the  frequency:  $v_0, \omega_0 $
on the other hand are linked to the 
NLS, while   the corresponding    multiplicative factors: $v_m, \omega_m $ to the SIT
contribution, 
as demonstrated  in   (\ref{V-0}). 
 We call $v(z)$ and $\omega(z) $ as the
velocity and the  modulation frequency for convenience, though strictly speaking their
derivatives $v^{'}(z)=v_0+v_m  N_0(z)$
 and $\omega^{'}(z)=\omega_0+\omega_m  N_0(z), $ 
should be the proper entries and be
called the {\it space}-velocity   and the {\it space}-modulation frequency,
respectively.  
  Consequently, the
   intensity of the optical field in the system
(\ref{NLSp},\ref{SITx}) given  as a solitonic beam  (\ref{1DNLSpsolit})
 can move   with a variable velocity 
 determined by the 
variable boundary   PI profile $ N_0(z) $.
Therefore the  {\it space}-acceleration of the  beam:
 $|v^{''}(z)|\equiv  {R^{-1}(z)}
=v_m  N^{'}_0(z) 
,$  becomes a  nontrivial  function 
  inducing  a bending of the  optical beam along a curve
with variable  curvature ${R(z)}$. 
It is illuminating to see, that for the conventional 
treatment with constant boundary PI giving   $N^{'}_0=0 ,$ the
 soliton acceleration would be zero inducing no 
bending for the  beam. However, 
 in the more general set up    
we are considering here, the intensity of the solitonic  beam (\ref{1DNLSpsolit})
 being   maximum  at $\xi(x,z)=0$,  the  optical beam
with its  peak intensity  would bend along the
curve $x=v(z)$ in the ($x, z$) plane,
 where  arbitrary function  $v(z)$ is
 fixed by the boundary setting of the
PI profile $N_0(z)$. Notice again, that  for   a
constant value of the   boundary 
PI: $N_0=1,$ the intensity peak of the beam reduces  to  a straight line
 $x=(v_0+v_m)z . $    
 Note, that unlike the
infinite energy Airy beam obtained earlier,
 the optical beam produced here is exponentially localized as a finite
energy soliton solution.  It is important to note for physical applications
that, since the curve-generating arbitrary function $v(z)$ is determined by
the variable boundary PI profile $N_0(z)$, one can tune the form of the
curve, along which the
 beam  would bend, by suitably designing and maintaining the PI
configuration of the dopant atoms at the transverse boundaries.

  We present below examples of 
 different curves (Fig.  1 a-d) along which the peak intensity of
the solitonic optical
 beam could  bend. 
These curves can be generated for  different
choices of the  arbitrary function $v(z)$,  
 linked to the transverse
boundary PI profile.
%   $N_0(z)=\frac 1 {v_m}(V^{'}(z)-v_0)$.

i)   $v(z)=a {z}^{ 2} ,$ would   induce   bending of an optical  beam
 along a {\it parabolic} curve (Fig.  1a).

ii)  $v(z)=(r-\alpha z^{ 2})^{\frac 1 2} $ corresponds to 
an {\it elliptic} curve along which an 
optical  solitonic  beam could bend (Fig. 1b). The curve naturally
  degenerates  to a {\it circle} 
%$ x^2+ z^2=1,$
 for $\alpha=1 $.

iii) $ v(z)=b {z}^{- 1 } $ would give  a  {\it
hyperbolic} curve for the bending of the  beam (Fig.  1c).

iv)  For the choice   $v(z)=a {\rm sin }k  z , \ $
 an optical  beam  would follow  
 a {\it periodic} curve  (Fig. 1d).

\begin{figure}[ht!] \begin{center}
\includegraphics[width=9.1cm,angle=0.0]{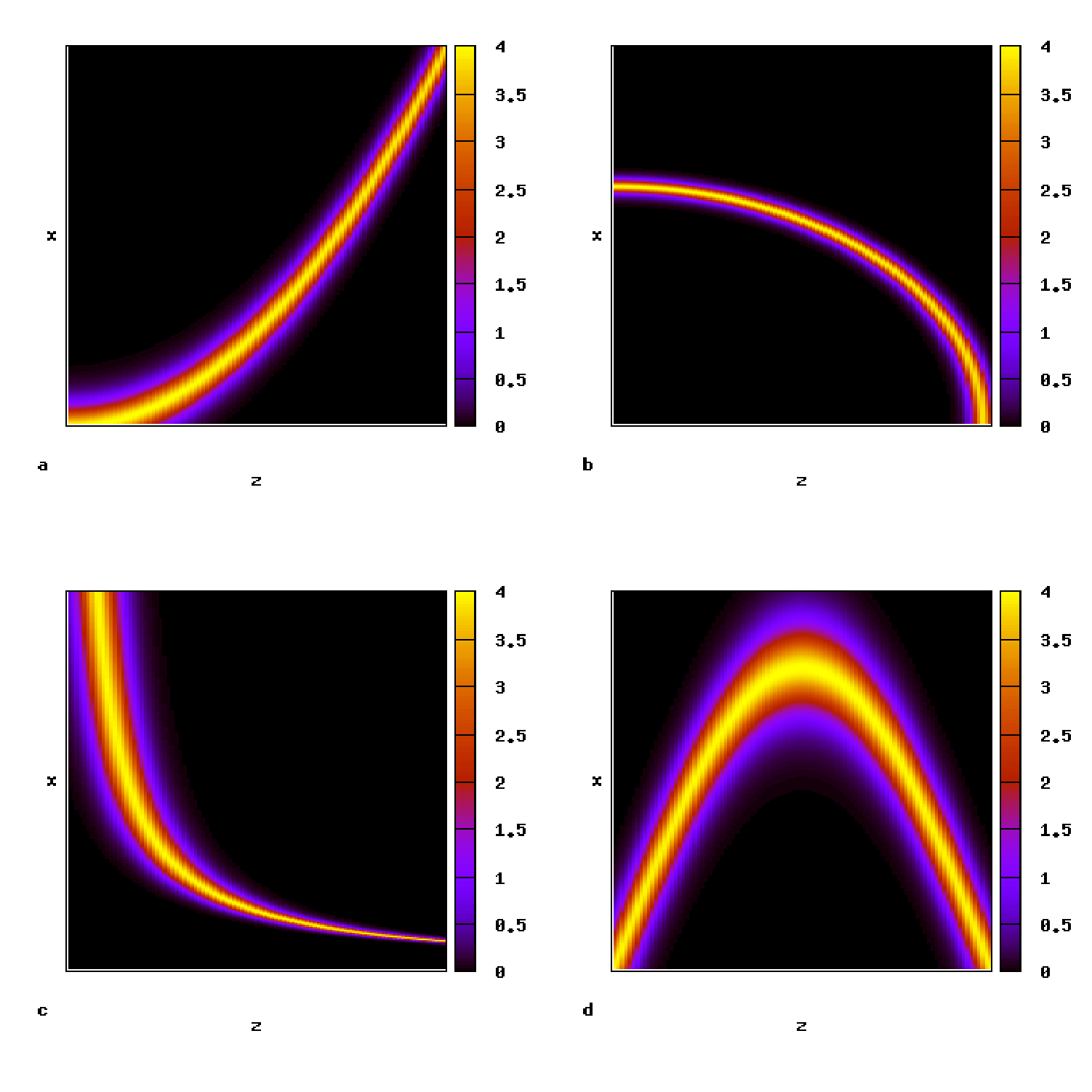}
\caption { Examples of the 
possible bending of an
% nonparaxial
 optical beam along different curves.
  The panels show the
peak intensity of an optical beam in the first quadrant of the plane
(a) along a parabolic
  curve $x= a z^2, \ a=0.1$, (b) an ellipse: $ x^2+ \alpha z^2=r,$ with
 $\alpha =0.4, \ r=40.0$,
(c)   a hyperbola   $xz=b, \ b= 8.0$ and
(d)  a periodic curve: $x=a{\rm sin }k z, $ with $a=8.0, \ k=0.3 .$   }
\end{center} \end{figure}
 This result demonstrates, that in our setup the bending of an
 % nonparaxial
optical 2D beam  can be achieved along any desired  planer curve,
 by  suitably  choosing  an arbitrary function,  linked   to
the boundary PI.  Therefore, the
nature of the curve along which the optical beam bends, can be
controlled by  presetting  the boundary  population inversion, which in turn
could be  realized   by
properly preparing the excitation profile of the dopant atoms along the
boundaries in the transverse dimension. Note, that  we obtain here
an exact self-accelerating   shape-invariant 
 solitonic beam in an integrable system,
 seemingly contradicting the common belief,
  that accelerating soliton can not exist in such a system 
   due to its strict conservation laws. 
Resolution  of this  apparent
paradox is that, here the acceleration of the soliton is linked to an
arbitrary  boundary function,
 which serves as an external source of energy and
should also be included for the validity of  the conservation laws. 

%\end{document}
 \section{ Arbitrary bending of 3D beam}
To generalize the above result on the   bending of an  optical beam  to
higher dimensions,  we include another transverse
    dimension $y$, since our bigger goal   is to achieve arbitrary 
 bending of a beam in a physical space.
Therefore we look for creating  an 
   optical soliton beam, that could move with 
  an arbitrary  acceleration in a  3D space.
Unfortunately, the conventional extension of the NLS equation to higher
dimensions with cubic nonlinearity
 will not  serve our purpose, since it  does not allow a 
stable soliton \cite{2Dnls}. We focus therefore 
   on  an integrable  
 extension  of the NLS equation obtained  
 by replacing the standard  cubic term  by a
current like nonlinearity \cite{arxiv12,2DiNLS}.
 In analogy with the  NLS-SIT system presented above,
 we  couple this  higher
dimensional  NLS equation:
 \begin{eqnarray} &&iE_z+E_{yx}+2iE(E^*E_x-
EE_x^*)=2p, \label{2DNLSp}
 \end{eqnarray}
to the SIT equation 
 (\ref{SITx}), where the interacting fields: the optical field,
the induced polarization  and the  PI of the dopant atoms, are 
given now in a 3D space, with $x,y $ as transverse
dimensions   and  $z$ as the optical axis.
    
Remarkably, the set of coupled higher-dimensional equations
(\ref{SITx})  and (\ref{2DNLSp})
turns out also to be an integrable system 
allowing a Lax pair formulation given by 
\begin{eqnarray}  U(\lambda)=U_2(\lambda), \  V(\lambda)= V_3(\lambda)+V_{-1}( \lambda), 
\label{UV2d}
\end{eqnarray}
where the  pair $( U_2(\lambda), \ V_3(\lambda))$ could be related to  the 
 integrable 2D NLS equation \cite{arxiv12,2DiNLS}
and $V_{-1}( \lambda) $ is given in the same form as   
(\ref{V-1}). The flatness condition (\ref{flat})
 of this Lax pair 
generates the integrable set of  coupled  2DNLS-SIT equations (\ref{2DNLSp},
\ref{SITx}) together with a 1D NLS like constraint. 
The  2DNLS-SIT equations
 admit also    solitonic beam solution for the optical field
with intensity:
 $|E|= 2{\rm sech}\xi, \ \xi= 2(x -vy-V(y,z)),$
moving with a self-accelerated motion in  space.
Since the  peak-intensity of the soliton  is  exponentially localized
at $\xi=0 $,
this   shape-invariant  optical beam 
 would self-bend over  an arbitrarily curved surface $x =vy+V(y,z) .$
Note that  the form of  this  surface embedded 
 in a 3D space is determined   by an arbitrary function
$V(y,z)$,  in  $y$ and $z$, which in turn is linked to  
the boundary  setting of the PI function along $x\to \pm \infty$:
  $ N(x \to \pm \infty,y,z)=N_0(y,z)= \frac 1 {v_m} (\partial_{z} V(y,z)),$ similar to the 2D case
(\ref{Vz}),
considered above.

Therefore, it is  possible to realize the bending of
 a 3D optical beam  over any preassigned surface
 by suitably maintaining  the population
inversion profile of dopant atoms along the  boundaries
in a  transverse direction $x$, generalizing the 2D model.
 We consider  an example of a
 surface: a circular paraboloid open along   $x$,
  over which a 3D beam can bend.  Fig.  2 shows a
 circular and a parabolic  curve obtained  as sections 
of this  surface.
 Such an optical  beam can curve around a 3D object
placed inside the hollow of the cup, turning it invisible.
 Other choices for the surface-generating function $V(y,z)$
 and hence for the boundary
 PI $N_0(y,z) $ would result to  other forms of
 the surface in a 3D space.
%\newpage 
%-------------
\begin{figure}[ht!]
\includegraphics[width=8.5cm,angle=0.0]{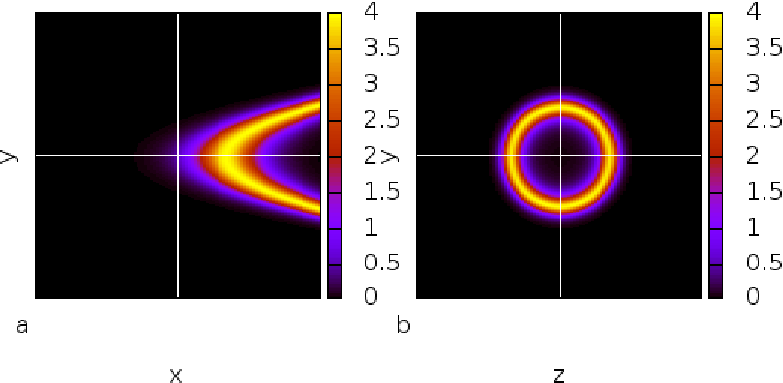}

\caption{ 
  The panels show  two sections of a paraboloid giving  
a) a parabolic curve  at a vertical section $z= 0.0$
and 
 b) a circle at   a horizontal section  $x=3.0$. 
Such a surface on which the peak intensity of  a 3D
 optical solitonic  beam is localized,  is 
obtained for a specific choice of the surface generating  function,
} 
 
%--------------------
\end{figure}

\section{Concluding remarks}
 
We have borrowed an idea, actively used in nonlinear optical fiber
communication, and applied it in the context of optical beam bending. 
The results obtained here show, that an optical beam
created in a nonlinear Kerr medium  in resonant interaction 
 with  an Erbium
doped medium,  can move
as a shape-invariant   
 localized soliton with a constant velocity. The constancy of the
soliton velocity is caused by 
  the conventional assumption of an   uniform
 boundary condition  for the population inversion  of the dopant atoms.
However, when the boundary values  of the population inversion
becomes nonuniform, the optical soliton  beam
 can acquire an arbitrary acceleration,
  inducing    the peak intensity of the beam to follow   an arbitrary
curve or a surface.  
 The desired form of this curved path traversed  by the beam, can be
controlled
 by carefully exciting the atoms of the doped medium,
 along the boundaries in
a transverse direction.  Excitation of the dopant atoms with variable
intensity, could be created and maintained at the boundaries by means of 
external  laser pumping.  This regulated  pumping of external energy into the
system would be  in fact responsible for the accelerated motion of the optical
solitonic beam. 
%_________________
The  outcome of the  present theoretical study is a
 possible prediction, viability of which would depend on the proper
 and clever experimental set up, which would be an experimental challenge.
%%___________  
 However in setting up the relevant experiment utmost care
should be taken in overcoming the difficulty faced in realizing the needed
stationary regime for the Bloch waves, since ultrafast fields are involved
in the SIT equations \cite{maimistov13}.  Nevertheless,
 we do hope that this theoretical  possibility of bending an optical beam
beyond the reported parabolic and circular curves, based on the
 idea used  in nonlinear  optical communication, would be feasible for
experimental verification,
 opening up new vistas of practical applications.  Unlike the infinite
energy spread-out solutions related to the Airy or Bessel functions used
earlier for obtaining the bending of an optical beam, our proposal based on
a finite energy soliton solution offers a strictly localized beam with a
 tightly curved path travelled by it.
  Moreover, due to the importance of the NLS equation and the soliton
solution in nonlinear phenomena in diverse areas in physics like plasma,
ocean wave, optical communication etc.  the possibility of creating
accelerating solitons in related systems should be of significant applicable
interest.  \section{Acknowledgement} A.  Kundu, as Senior Associate of ICTP,
Italy wants to thank the Centre,
  where this work was
completed, for its continuing support. 

\end{document}